\documentclass[review]{elsarticle}

\usepackage{hyperref}
\usepackage{lineno}
\usepackage{multirow}
\usepackage{multicol}
\usepackage{gensymb}
\usepackage{amssymb}
\usepackage{subcaption}
\usepackage{tablefootnote}
\usepackage{caption}
\usepackage{subcaption}

\newcommand{\neutron}{n$_\mathrm{eq}$/cm$^{2}$}
\newcommand{\mum}{~$\mathrm{\mu m}$}

\newcommand{\fluence}[2]{$#1\cdot10^{#2}$~n$_\mathrm{eq}$/cm$^{2}$}

\journal{NIM Section A}

\begin{document}
\begin{frontmatter}

\title{Optimization of the Gain Layer Design of Ultra-Fast Silicon Detectors}

\author[address1,address3]{F. Siviero\corref{mycorrespondingauthor}}
\cortext[mycorrespondingauthor]{Corresponding author}
\ead{federico.siviero@edu.unito.it}
\author[address4]{R. Arcidiacono}
\author[address2]{G. Borghi}
\author[address2]{M. Boscardin}
\author[address3]{N. Cartiglia}
\author[address2]{M. Centis Vignali}
\author[address1]{M. Costa}
\author[address5]{G-F. Dalla Betta}
\author[address4]{M. Ferrero}
\author[address2]{F. Ficorella}
\author[address1]{G. Gioachin}
\author[address3]{M. Mandurrino}
\author[address6,address7]{S. Mazza}
\author[address3,address1]{L. Menzio}
\author[address5]{L. Pancheri}
\author[address2]{G. Paternoster}
\author[address6,address7]{H-F W. Sadrozinski}
\author[address6,address7]{A. Seiden}
\author[address3]{V. Sola}
\author[address1]{M. Tornago}

\address[address1]{Università degli Studi di Torino, Torino, Italy}
\address[address2]{Fondazione Bruno Kessler, Trento, Italy}
\address[address3]{INFN, Torino, Italy}
\address[address4]{Università del Piemonte Orientale, Novara, Italy}
\address[address5]{Università degli Studi di Trento, Trento, Italy}
\address[address6]{University of California, Santa Cruz, USA}
\address[address7]{Santa Cruz Institute for Particle Physics, USA}

%\linenumbers

\begin{abstract}
%\linenumbers
In the past few years,  the need of measuring accurately the spatial and temporal coordinates of the particles generated in  high-energy physics experiments has spurred a strong R\&D in the field of silicon sensors. Within these research activities, the so-called Ultra-Fast Silicon Detectors (UFSDs), silicon sensors optimized for timing based on the Low-Gain Avalanche Diode (LGAD) design, have been proposed and adopted by the CMS and ATLAS collaborations for their respective timing layers. The defining feature of the Ultra-Fast Silicon Detectors (UFSDs) is the internal multiplication mechanism, determined by the gain layer design. In this paper, the performances of several types of gain layers, measured with a telescope instrumented with a $^{90}$Sr $\beta$-source, are reported and compared. The measured sensors are produced by Fondazione Bruno Kessler (FBK) and Hamamatsu Photonics (HPK). 

% A new figure of merit has been introduced to assess the radiation hardness of the different gain layer designs tested: the voltage increase required to provide 10~fC after a certain irradiation fluence with respect to the pre-irradiation condition. Thanks to this figure, we demonstrated that the innovative deep gain layer design is more resistant to radiations than the standard gain implant. 

The sensor yielding the best performance, both when new and irradiated, is an FBK 45\mum-thick sensor with a carbonated deep gain implant, where the carbon and the boron implants are annealed concurrently with a low thermal load. This sensor is able to achieve a time resolution of 40~ps up to a radiation fluence of~\fluence{2.5}{15}, delivering at least 5~fC of charge. 

%The FBK sensors feature three different gain layer designs, with boron and carbon co-implantation, and three sensor thicknesses (35, 45, and 55\mum), while the HPK sensors have two different gain layer designs (not carbonated) and two thicknesses (45 and 80\mum). 

%We also introduce a new figure of merit to assess the radiation hardness of the UFSD gain layer design: the voltage increase required to provide 10~fC after a certain irradiation fluence with respect to the pre-irradiation condition; thanks to this figure, 

\end{abstract}

\begin{keyword}
UFSD \sep LGAD \sep radiation resistance \sep time resolution
\MSC[2010] 00-01\sep  99-00
\end{keyword}

\end{frontmatter}

\tableofcontents

%\linenumbers

\section{Introduction}

Silicon sensors are presently the detector of choice in high-precision timing systems for several future high-energy physics (HEP) experiments. The requirements on the time resolution depends on the specific experiment: according to the Detector R\&D roadmap of the European Committee for Future Accelerators (ECFA)~\cite{ECFA}, the time resolution for the detectors at the High-Luminosity LHC (HL-LHC) is required to be of about 30-40~ps, 20-30~ps at the Electron-Ion Collider (EIC), and it decreases to about 10~ps for the Time-of-Flight (TOF) systems at the Future Circular Collider (FCC), and other Higgs-Electroweak-Top factories.

The present state-of-the-art designs of the high-precision timing systems are the timing layers of the CMS~\cite{CMS_TDR} and ATLAS~\cite{ATLAS_TDR} collaborations, which are being developed for the HL-LHC. The endcap regions of such timing layers, in particular, will be instrumented with thin ($\sim$~50~\mum-thick) Low Gain Avalanche Diode (LGAD~\cite{PELLEGRINI201412}) sensors optimized for timing, also known as Ultra-Fast Silicon Detectors (UFSDs~\cite{SADROZINSKI201618}). UFSDs combine an optimized sensor geometry with a moderate internal gain to achieve a time resolution of 30-40~ps for minimum ionizing particles (MIPs): a thorough description can be found in~\cite{MF_book}. 

The timing performance is not the only aspect to be considered when designing a timing system: the radiation hardness is key, but also the material budget, the efficiency, and the possibility of building large-area detectors play an important role. UFSDs are presently one of the most promising detectors for high-precision timing because they address these requirements as well:

\begin{itemize}
    \item They are radiation hard, able to withstand radiation fluences up to about \fluence{1}{15}~without loss  of performance~\cite{FERRERO201916}.
    \item They have a reduced material budget, given their reduced thickness.
    \item Silicon is 100\% efficient when detecting the passage of an ionizing particle.
    \item They can be manufactured in large quantities, as standard silicon sensors. For example, the CMS Endcap Timing Layer~\cite{CMS_TDR} will use~$\sim$~14 m$^{2}$ of UFSDs.
\end{itemize}
 
In order to optimize the UFSDs design, in this work we measured and compared the performances of five different types of gain layer and four different active sensor thicknesses using a dedicated $\beta$-source setup. The tested devices have been manufactured by Fondazione Bruno Kessler (FBK)~\cite{fbk_site} and Hamamatsu Photonics (HPK)~\cite{hpk_site}. Almost all sensors have been irradiated, without bias, with neutrons at the JSI TRIGA research reactor in Ljubljana \cite{SNOJ2012483}, up to a fluence of $2.5\cdot10^{15}$~\neutron~(in the following, the radiation fluence will be sometimes identified with a $\Phi$ as a short-hand notation).

The measurement campaign offered the possibility of comparing an unprecedented variety of gain layer designs.  To assess their relative  radiation resistance,  we introduced the figure of merit $\Delta$V$_{10fC}$, the voltage increase as a function of fluence  needed to deliver a signal of 10 fC. This quantity not only capture the effect of acceptor removal,  normally expressed in terms  of the acceptor removal coefficient $c$~\cite{FERRERO201916},  but also the recovering capability of a bias increase, which depends on the gain implant position.  

Thanks to this new quantity, we demonstrate that deep gain implants, despite having a slightly larger $c$, are more radiation resistant as they have a stronger gain recovery capability with bias than the standard gain implants.  

The introduction of $\Delta$V$_{10fC}$ has been instrumental to achieve the most important result of this work: the identification of the best gain layer design among those tested, which is that of the FBK UFSD3.2 W13, a 45\mum-thick sensor with a carbonated deep gain layer where the carbon and the boron implants are annealed concurrently with a low thermal load. 

%Thanks to this result, FBK chose the UFSD3.2 W13 as reference design for the UFSDs with deep implant of its latest production, UFSD4~\cite{MT_UFSD4}.

\section{FBK UFSD3.2 production} 
\label{sec:ufsd32}
\begin{figure}[h]
    \centering
    \includegraphics[width=\linewidth]{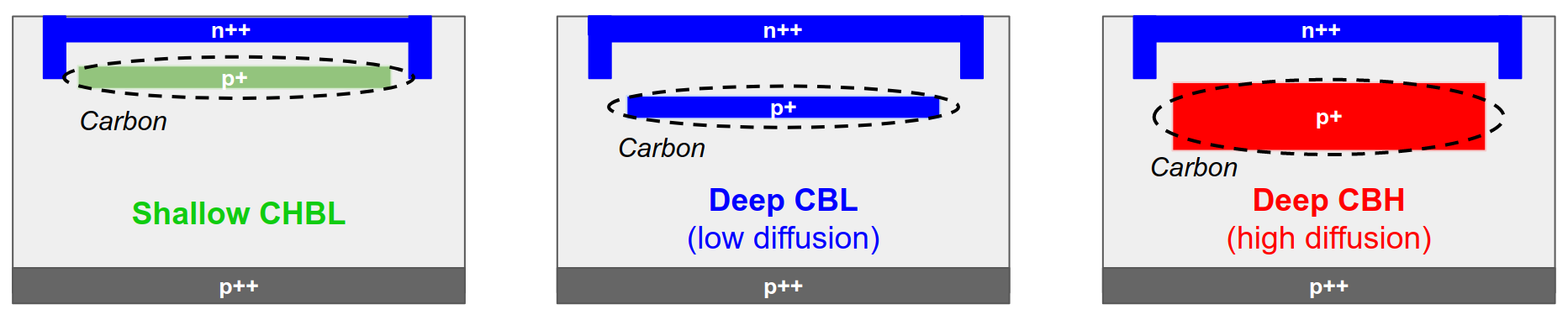}
    \caption{The different types of gain implants of the UFSD3.2 production.}
    \label{fig:gain_layer_types}
\end{figure}

One of the main goals of the FBK UFSD3.2 production is the investigation of different gain layer designs. In this production, two different depths of the gain implant, shallow and deep~\cite{CARTIGLIA-34thRD50}, were used. The gain layer implants have been co-implanted with carbon, to enhance their radiation resistance~\cite{FERRERO201916}. In order to study the interplay between carbon and boron implants, three different doping activation cycles were used: (i) CHBL (Carbon-High, Boron-Low), meaning that carbon is activated with an high thermal load while boron is activated with a low thermal load; (ii) CBL (Carbon-Boron Low), when carbon and boron implants are activated together at low thermal load; (iii) CBH (Carbon-Boron High), when carbon and boron implants are activated together at high thermal load. CHBL is used for shallow gain implants, while CBL and CBH for deep implants. Figure~\ref{fig:gain_layer_types} shows the different gain layer types. The sketch on the left illustrates the gain layer with a shallow implant, while the center and right sketches those with deep implants.  

Prior to UFSD3.2, FBK only used shallow gain implants: in this geometry, the high electric field region is narrower than 1\mum; whereas, for the deep implants, the high-field region extends for about 1.5-2\mum. 

The gain implants, shallow or deep, feature five different doses of boron and four of carbon. A summary of the main characteristics of the wafers can be found in table \ref{tab:UFSD3.2}, where the boron and carbon doses are reported in arbitrary units.  The production also features the first 35\mum-thick sensors produced by FBK, which have been measured only un-irradiated.

\begin{table}[htb]
    \centering
    \tabcolsep3pt
    \begin{tabular}{cccccc}
    \hline
        Wafer & Thickness [\mum] & GL depth & B dose [a.u.] & C dose [a.u.] & Diffusion scheme\\
         \hline\hline
         1 & 45 & \multirow{4}{*}{Shallow} & 0.98 & 1 & \multirow{4}{*}{CHBL}\\
         3 & 45 & & 0.98 & 0.8 & \\
         6 & 35 & & 0.94 & 1 & \\
         7 & 55 & & 0.98 & 1 & \\
         \hline
         10 & 45 & \multirow{3}{*}{Deep} & 0.70 & 0.6 & 
         \multirow{3}{*}{CBL}\\
         12 & 45 & & 0.74 & 1 & \\
         13 & 45 & & 0.74 & 0.6 & \\
         \hline
         14 & 45 & \multirow{4}{*}{Deep} & 0.74 & 1 & \multirow{4}{*}{CBH}\\
         15 & 55 & & 0.74 & 1 & \\
         18 & 45 & & 0.78 & 1 & \\
         19 & 45 & & 0.78 & 0.6 & \\[0.5ex]
    \hline
    \end{tabular}
    \caption{The wafers of the UFSD3.2 production tested in this work. The active thickness, the gain layer (GL) depth, the boron (B) and carbon (C) doses and the diffusion scheme are reported.  }
    \label{tab:UFSD3.2}
\end{table}

%Figure \ref{fig:pin-lgad} shows an example of a UFSD3.2 2x1 array used in this testing campaign: one pad is an LGAD and one is a standard PIN diode with no gain implant

%\begin{figure}
%    \centering
%    \includegraphics[width=0.5\linewidth]{Figures/pin-lgad.png}
%    \caption{UFSD3.2 2X1 array.}
%    \label{fig:pin-lgad}
%\end{figure}

\section{Hamamatsu  HPK2 and 80D productions.}

 Two different types of sensors manufactured by Hamamatsu Photonics (HPK) are included in this analysis. The first type belongs to the HPK2 production~\cite{MF_book}: it has a  non-carbonated deep gain implant and a 45\mum~active thickness. The HPK2 production has four gain implant doping levels: in this study, only the least doped sensors were used, the so-called HPK2 split 4. The second type of sensors here analysed belongs to an earlier HPK production, called HPK 80D, which features 80\mum~thick sensors with a shallow gain implant. The HPK 80D sensors have been measured un-irradiated only.

\section{The $\beta$-source experimental setup}\label{sec:ExpSetup}
The $\beta$-source setup is used to study the characteristics of the signals induced by charged particles (in this case low energy electrons) in the UFSD, in order to assess the performance of the device under test (DUT).

The DUTs are mounted on a 10 $\times$ 10 cm$^{2}$ read-out board designed at the University of California Santa-Cruz (it will be called SC board hereafter) \cite{CARTIGLIA201783, beta-sc}. The SC board has a fast inverting amplifier with a trans-impedance of about 470~$\Omega$, followed by a 20 dB broadband amplifier (a 2 GHz Cividec model C1HV), for a total trans-impedance of 4700 $\Omega$. A Teledyne-Lecroy HDO9404 oscilloscope with a 20 GS/s sampling rate, corresponding to a 50~ps time discretization, and 10 bit vertical resolution is used to record the signals.

\begin{figure}[h]
    \centering
    \includegraphics[width=0.75\linewidth]{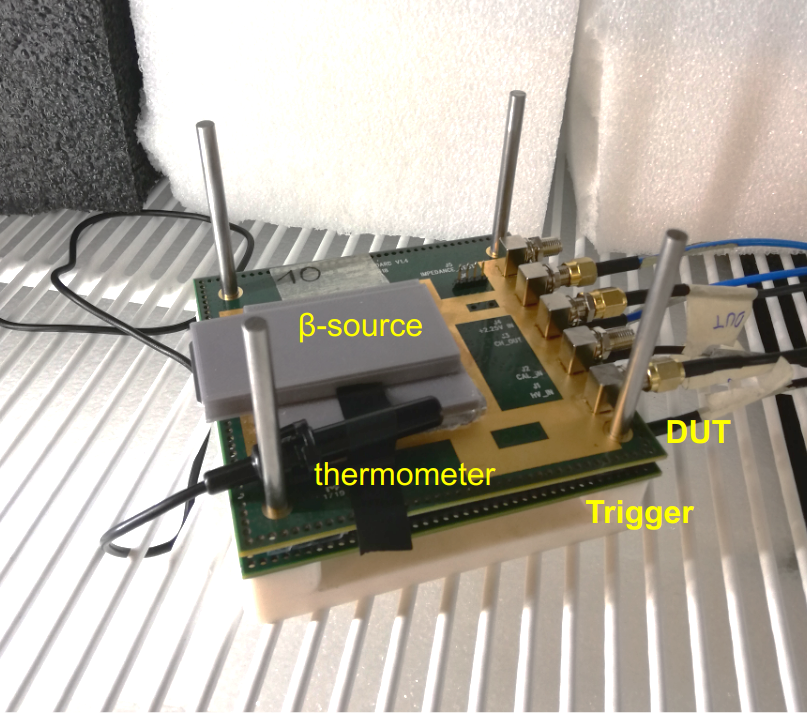}
    \caption{The 3D-printed telescope used during the measurement campaign. The case containing the $\beta$-source and the thermometer are also indicated.}
    \label{fig:telescope}
\end{figure}

The DUT and a trigger plane, whose characteristics are described in the following section, are placed in a 3D-printed structure (figure~\ref{fig:telescope}) for alignment, while the $^{90}$Sr $\beta$-source with an activity of 3.6 kBq is inserted in a second \\ 3D-printed case. The $\beta$-source is then positioned on top of the telescope formed by the DUT and the trigger.

Electrons from a $^{90}$Sr source have two energy branches: one with low energy from $^{90}$Sr ($< 0.5$ MeV), and one with higher energy from $^{90}$Y (0.5 - 2 MeV). The low-energy electrons do not manage to traverse hundreds of micron of silicon (the total UFSD thickness, given by the active region plus the support wafer underneath, is in the 300-600~\mum~range, depending on the producer), and get stopped in the first sensor of the telescope. The DUT board is placed on top of the stack for this reason, so that only electrons from the high-energy branch, which are mostly MIPs, can reach the trigger. 

The coincidence rate due to the $\beta$-source is enhanced by opening a hole in the SC board aligned with the DUT active region. 

The telescope is placed inside a climate chamber in order to perform measurements at -25 $\degree$C and with a relative humidity of less than 10\%. Both temperature and humidity are  monitored with a thermo-hygrometer placed close to the sensor and found to be very stable.

\subsection{Trigger characterization}
The trigger used in the data acquisitions here analysed is a 50\mum~thick single-pad HPK UFSD  with an area of 1~$\times$~3 mm$^{2}$. Its time resolution was measured using two identical devices and dividing the measured total resolution by $\sqrt{2}$. A resolution of $\sigma_{trigger}$ = 31.6 $\pm$ 1.3~ps has been measured when the sensor is biased at 175~V and operated at -25 $\degree$C, which are its standard running conditions in this setup.

During the measurements presented in this work, the trigger threshold was set to 30~mV, more than 10 times its RMS noise level. The trigger rate is approximately 2 Hz, stable during all measurements, with a 10\% of in-time signal coincidences between the DUT and the trigger. Roughly 2000 signal coincidences are recorded for each bias step, requiring approximately 20k triggers.

\section{Description of the Data Analysis Methods}
Several important parameters related to the sensor performance can be measured using the $\beta$-source setup previously described: the collected charge; the electronic noise (sensor-amplifier dependent);  the value of the gain; the time of arrival (ToA) and the time resolution.

%The signal waveforms of the trigger and the DUT were recorded with the oscilloscope during data taking. 
For all these measurements, the offline event selection requires: (i) the trigger signal amplitudes to be in the 80-250~mV range; (ii) the trigger collected charge in the 11.5-53.0~fC range; (iii) the DUT signal amplitude above 10~mV, about 5 times the RMS noise; (iv) the signal not to saturate either the oscilloscope vertical scale or the amplifier. Point (iv) rejects events with DUT signal amplitudes greater than 300~mV. 

\begin{figure}[h]  
\centering 
\includegraphics[width=\textwidth]{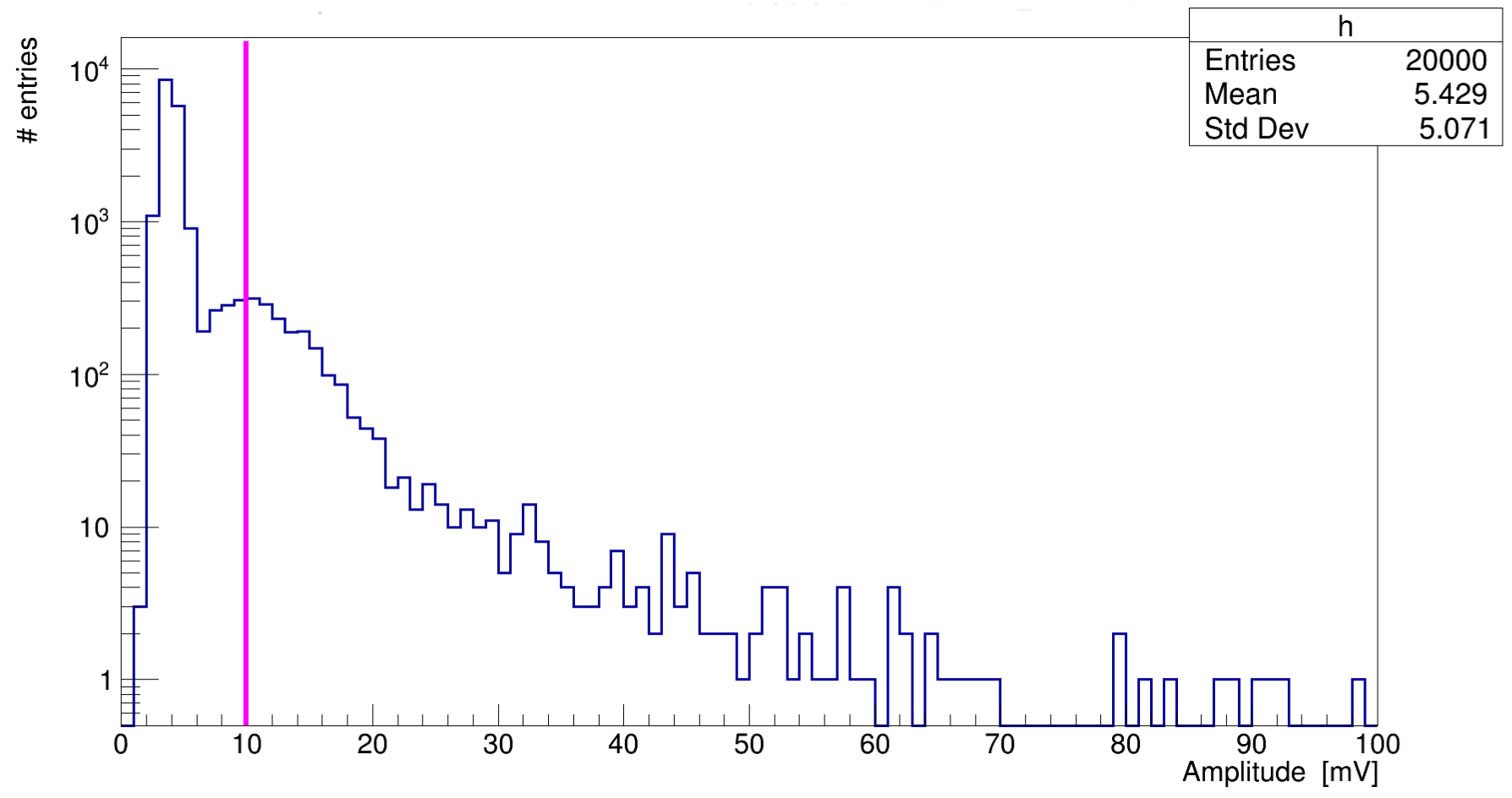}
\caption{Amplitude distribution of the FBK W14 irradiated at a fluence $\Phi$~=~$2.5\cdot10^{15}$~\neutron, biased at 500~V. The pink line shows the amplitude cut applied in the event selection. Measurement performed at -25 $\degree$C.}
\label{fig:w14_irrad}
\end{figure}

Point (iii) introduces a small bias in the analysis of the sensors irradiated at a fluence $\Phi = 2.5\cdot10^{15}$~\neutron~when biased at voltages $\leq$ 500 V.  Figure~\ref{fig:w14_irrad} shows the Landau distribution  of an  FBK UFSD3.2 W14 sensor, which is the least radiation hard, irradiated at $2.5\cdot10^{15}$~\neutron, and biased at 500~V.  In this case, the lower tail of the Landau distribution partially overlaps with the noise.

\subsection{Collected Charge, Noise, and Gain}
The measurement of the collected charge for one event is derived from the signal area, obtained by integrating the signal over time using the Simpson's method \cite{simpson}, divided by the read-out system trans-impedance,  4700 $\Omega$. A baseline subtraction is performed on each waveform, using the first 100 sampled points (5~ns), taken before the signal occurs. The same 100 points are also used to measure the sensor-electronics RMS noise.

The gain of the UFSD sensor can be defined as the ratio of the charge collected by the UFSD at a given voltage, $Q(V)$, over that collected by an equivalent PIN diode, $Q_{PIN}$, at a voltage below 500~V, so that gain in the bulk cannot happen (see \cite{MF_book} for details on gain in the UFSD bulk):

\begin{equation}
    Gain = \frac{Q(V)}{Q_{PIN}}
\label{eq:gain}    
\end{equation}

where $Q(V)$ and $Q_{PIN}$ refer to the most-probable-value (MPV) of the UFSD and PIN collected charge distributions, respectively. The definition of equation~\ref{eq:gain} is such that it considers both the UFSD gain from the gain layer and that from the bulk. Since $Q_{PIN}$ is always assessed at a voltage below 500~V and the gain measurements are performed at fixed temperature, it can be considered a constant.

Since experimentally the measurement of the charge collected by the PIN diode leads to large uncertainties (given its rather small value, about 0.5~fC), it was decided to take $Q_{PIN}$ from theory (assuming to compute it at a voltage $<$~500): $\sim$~0.45~fC in a 45\mum~thick sensor and $\sim$~0.55~fC in a 55\mum~one~\cite{Passeri}.  The charge computed from theory is affected by the uncertainty on the active sensor thickness, which results in a relative uncertainty of $\sim$~4\% for a 50\mum-thick sensor, much lower than the one affecting the $Q_{PIN}$ measurement ($\ge$~20\%). 

The theoretical $Q_{PIN}$ value has been used to compute the gain of UFSDs either before and after irradiation.

\subsection{Time of Arrival and Time Resolution} 

The time of arrival (ToA) of a signal is computed using the constant-fraction discriminator (CFD) method, i.e., it is defined as the time at which the signal crosses a certain fraction of the total signal amplitude. A 20\% threshold (CFD20) provided the best results for most of the sensors tested. This fraction was increased to 30\% (CFD30) for a few heavily irradiated sensors due to their higher noise and a slightly different signal shape. 

For the evaluation of the time resolution of a sensor biased at a given value, the distribution of the difference between the DUT ToA  and the trigger ToA is plotted, and then fitted with a Gaussian function whose standard deviation is the squared sum of the DUT and trigger resolution, hence:

\begin{equation}
    \sigma_{DUT} = \sqrt{\sigma_{measured}^{2}-\sigma_{trigger}^{2}}
\end{equation}

\subsection{Evaluation of the uncertainties on the measured quantities} % systematic?
Due to the large number of sensors tested and the relatively low trigger rate, it would have been very challenging to assess the uncertainties on the measured quantities by repeating the measurement of each sensor several times. However, several steps were taken to keep the uncertainties under control: (i) the same trigger has been used in each of the 250 independent measurements of this campaign, (ii) the trigger plane was always operated at the same temperature, and voltage; and (iii) for a given sensor (pre-irradiation UFSD3.2 W7), a set of repeated measurements was carried out to evaluate all the uncertainties. 

Table~\ref{tab:uncertainties} shows the list of uncertainties affecting the quantities measured at the $\beta$-source setup.

\begin{table}[htbp]
\centering
\caption{\label{tab:uncertainties} The uncertainties on the main quantities measured at the $\beta$-source setup.}
\smallskip
\begin{tabular}{c|c}
 & uncertainty \\
\hline
time resolution & 1.5~ps  \\
collected charge & 0.2~fC\\
gain & 0.5\\
RMS noise & 0.2~mV \\
Slew rate & 2.9~mV/ns  \\ \hline
\end{tabular}
\end{table}

\section{Characterization of a gain layer design:\\ $Q, \sigma_t$, V$_{10fC}$, $\Delta$V$_{10fC}$($\Phi$), $\alpha$, and noise }

The evaluation of a gain layer design needs to include several aspects.  Obviously, the time resolution of the sensor is the first quantity to be considered in its characterization.  However, the time resolution achieved in laboratory tests with a state-of-the-art analog read-out board (as the one used in this work) does not guarantee that the same results can be achieved with much more complex low-power electronics at the experiments, which usually feature a higher electronic noise. 

Additional quantities, such as the collected charge and the noise level, need to be evaluated, as they determine the final achievable performance. Considering the requirements of future front-end ASICs, for example~\cite{ETROC}, a signal charge above 8-10~fC helps achieving a time resolution $\sigma_t < 50$~ps. In general, the higher the charge, the better the front-end ASIC performs.  Likewise, the noise arising from the UFSD gain  mechanism (see~\cite{ROPP}) needs to be kept as low as possible, below the front-end electronic noise. 

Another important aspect of a design is its ease of use: to operate UFSDs reliably, the charge-bias characteristics ($Q(V)$) must not be too steep. A very steep curve has two significant consequences: (i) it amplifies the effect of gain implant non-uniformity, as a small doping variation yields to very large gain differences, and (ii) a small change in biasing condition leads to very large gain variations. 

The steepness of the charge curve is quantified by fitting the charge-bias characteristics with an exponential fit:

\begin{equation}
    Q(V) = Q_{PIN} \cdot e^{\alpha \cdot V}
\label{eq:gain}
\end{equation}

with the $\alpha$ parameter being a measurement of the curve slope; $Q_{PIN}$ was previously introduced and represents the charge collected by a PIN diode in absence of bulk gain.

In the evaluation of gain layer designs, it is also important to consider at what bias voltage a given charge is reached. In the present analysis, the charge Q =  10~fC was used as a bench point and the value V$_{10fC}$ was measured for every sensor.  

Lastly, it is also important to consider the evolution of the bias point V$_{10fC}$ with irradiation. For this reason, the quantity $\Delta$V$_{10fC}$($\Phi$) is introduced. It describes the voltage increase required to provide 10~fC after a certain irradiation fluence $\Phi$, with respect to the pre-irradiation condition.  Such voltage increase is needed to compensate for the gain decrease, due to the acceptor removal mechanism, as extensively reported in \cite{MF_book}. $\Delta$V$_{10fC}$($\Phi$) is defined as:

\begin{equation}
    \Delta V_{10fC}(\Phi) = V_{10fC}(\Phi) - V_{10fC}(\Phi=0)
    \label{eq:DV}
\end{equation}

where V$_{10fC}(\Phi)$ is the bias required to provide 10~fC after an irradiation fluence $\Phi$, and V$_{10fC}(\Phi=0)$ is the bias required to provide 10~fC before irradiation.

A sensor with a small $\Delta V_\Phi$(10~fC) has two key advantages: (i) it can sustain higher fluences before reaching the breakdown condition (i.e. it is more radiation resistant), and (ii) it delivers more uniform response when exposed to non-uniform irradiation.

\section{Experimental Results}

\subsection{Collected charge $Q$}

\begin{figure}[h]
\centering
\includegraphics[width=\textwidth]{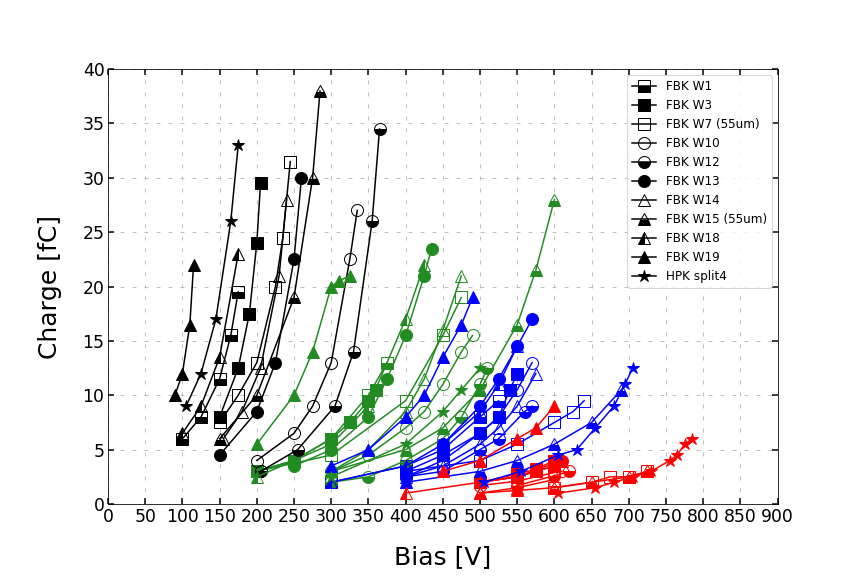}
\caption{Collected charge as a function of bias voltage. Sensors irradiated at different fluences are shown with different colors (black = pre-rad, green = $8\cdot10^{14}$~\neutron, blue = $1.5\cdot10^{15}$~\neutron, red =  $2.5\cdot10^{15}$~\neutron). Measurements performed at -25 $\degree$C.}
\label{fig:charge}
\end{figure}

Figure~\ref{fig:charge} shows the collected charge as a function of bias for DUTs from 10 wafers of the FBK UFSD3.2 production and from the HPK2 split 4. The color-code of the plot, common to other similar plots in this report,  identifies different irradiation fluences: pre-rad sensors are in black, sensors irradiated at a fluence of~ $8\cdot10^{14}$~\neutron~are in green, $1.5\cdot10^{15}$~\neutron~ in blue, $2.5\cdot10^{15}$~\neutron~ in red. 

The trend of the collected charge versus  irradiation is as expected: as the fluence increases, the bias voltage needs to be raised to compensate for the acceptor removal effect. All tested sensors are able to deliver at least 10~fC  up to a fluence of $1.5\cdot10^{15}$~\neutron. At the highest fluence, $2.5\cdot10^{15}$~\neutron, the gain is too low even at high bias values, and none of the sensors deliver 10~fC. FBK UFSD3.2 W19 is the best performer at high fluence, delivering 9~fC at bias V = 600~V. 

For each sensor, the point at the highest bias has been taken at the maximum achievable stable biasing condition: at bias voltages above the reported measurements, either the noise is too large, or the sensor goes into a breakdown.  

\subsection{Time resolution $\sigma_t$}

\begin{figure}[h]
\centering 
\includegraphics[width=\textwidth]{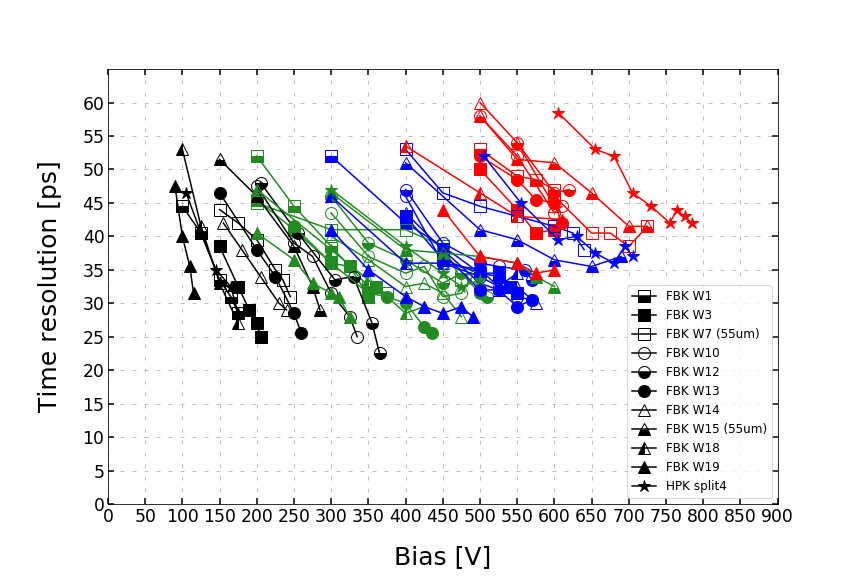}
\caption{Time resolution as a function of bias voltage. Sensors irradiated at different fluences are shown with different colors (black = pre-rad, green = $8\cdot10^{14}$~\neutron, blue = $1.5\cdot10^{15}$~\neutron, red =  $2.5\cdot10^{15}$~\neutron). Measurements performed at -25 $\degree$C.}
\label{fig:res_vs_bias}
\end{figure}

Figure~\ref{fig:res_vs_bias} shows the time resolution corresponding to the collected charges points shown in figure~\ref{fig:charge}. All sensors achieve a resolution $\sigma_t <$  40~ps up to a fluence of $1.5\cdot10^{15}$~\neutron; at $2.5\cdot10^{15}$~\neutron~the resolution degrades up to $\sigma_t \sim$ 45~ps. It is worth pointing out that the very good result achieved at the highest fluence is obtained despite the rather small signal charge, $Q~\leq~5~fC$, thanks to the characteristics of the read-out board used in this work.

\subsection{V$_{10fC}$, $\Delta$V$_{10fC}$($\Phi$), and $\alpha$.}
The collected charge as a function of bias ($Q(V)$) for each pre-rad sensor (black points in figure~\ref{fig:charge}) has been used to extract the slope parameter $\alpha$ and V$_{10fC}$. Table \ref{tab:slopes} reports the results of the fits of each curve with equation~\ref{eq:gain}. 

The gain implant doping level controls both parameters:  the higher the level, the larger the slope $\alpha$ and the smaller the V$_{10fC}$ value. A small value of V$_{10fC}$ is beneficial, as it leads to lower power consumption; however, it also implies a very steep gain curve at the start of the experiment.

\begin{table}[h]
\caption{\label{tab:slopes} Slopes $\alpha$ [$mV^{-1}$] of the $Q(V)$ curves and V$_{10fC}$[V] of the tested sensors.}
\begin{tabular}{|c|c|c|c|c|c|c|c|c|c|c|c|}
\hline
Wafer & 1 & 3&7&10&12&13&14&15&18&19& HPKs4 \\
\hline
$\alpha$  & 17 & 16 & 14  & 9 & 9 
                     & 12 & 13 & 12 & 18 & 26 & 20 \\
\hline
V$_{10fC}$  & 140 & 165 & 143 & 280 & 310
             & 210 & 190 & 164 & 130 & 90 & 110 \\
             \hline
\end{tabular}
\end{table}

\begin{figure}[h]
\centering 
\includegraphics[width=\textwidth]{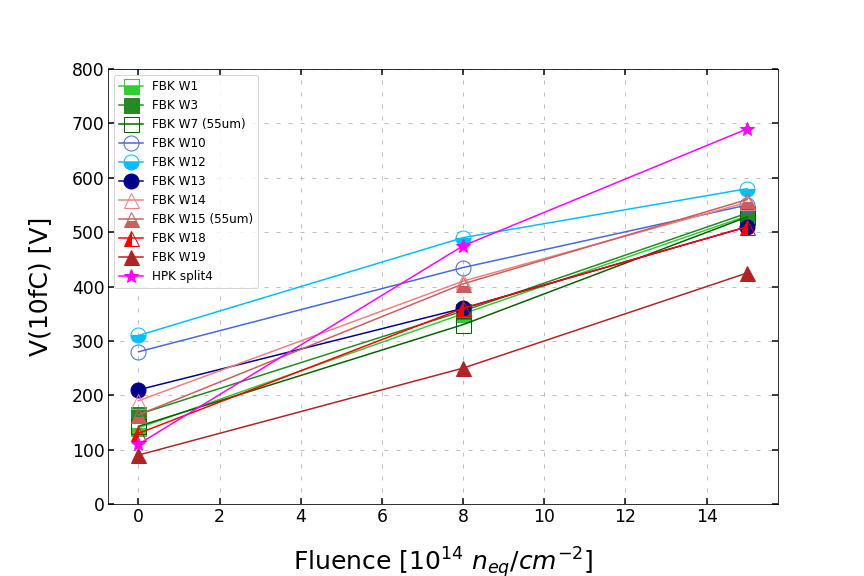}
\caption{The parameter V$_{10fC}$ shown as a function of the irradiation fluence level. Values extracted from measurements performed at -25 $\degree$C.}
\label{fig:V(10fC)}
\end{figure}

Figure \ref{fig:V(10fC)} presents the parameter V$_{10fC}$ as a function of the irradiation fluence. The voltages of the two 55\mum-thick sensors (W7 and W15) are scaled by a factor $\frac{45}{55}$, so that they can be compared with the 45\mum-thick ones.

Pre-irradiation sensors show a large spread in  V$_{10fC}$, ranging from $\sim$~100~V to more than 300~V; the spread is maintained at a fluence $8\cdot10^{14}$~\neutron, with sensors requiring between 250 and 500~V to provide 10~fC. At fluence $1.5\cdot10^{15}$~\neutron,  the majority of sensors reach the target working point between 500 and 600~V.

\begin{figure}[h]
\centering 
\includegraphics[width=\textwidth]{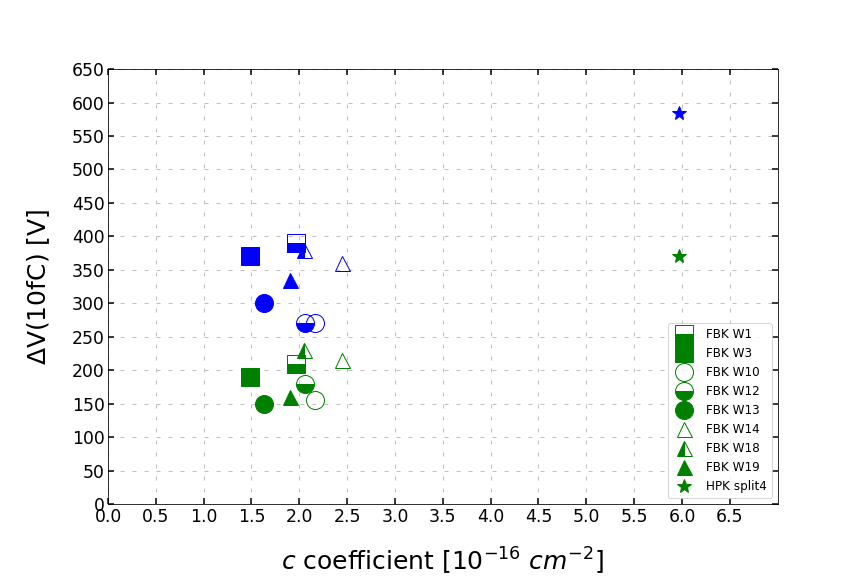}
\caption{$\Delta$V$_{10fC}$($\Phi$) shown as a function of the acceptor removal coefficient \textit{c}. Green markers represent $\Delta$V calculated after a fluence of $8\cdot10^{14}$~\neutron, whereas blue ones are calculated at \fluence{1.5}{15}. This plot reports results only from 45\mum-thick sensors. Values extracted from measurements performed at -25 $\degree$C.}
\label{fig:DV}
\end{figure}

The spread in V$_{10fC}$ decreases with increasing fluence since, at high fluences, only a small fraction of the gain implant survives and all the sensors have roughly the same gain at a given voltage, despite their different gain layer designs. Overall, small differences in  gain layer design become gradually less important with irradiation. Two notable exceptions are shown in figure~\ref{fig:V(10fC)}: (i) UFSD3.2 W19, due to its design, maintains at high fluences a large fraction of gain implant so V$_{10fC}$ is reached at $\sim$~400~V; (ii) HPK split 4, despite starting with a low value of V$_{10fC}$, has a low radiation resistance (it has the highest acceptor removal coefficient \footnote{\textit{c} describes the exponential removal of the gain implant N with fluence $N(\Phi) = N(\Phi = 0)*e^{-c*\Phi}$~\cite{MF_book, ROPP}}, see figure~\ref{fig:DV}) and requires 700~V to provide 10 fC at $\Phi =  1.5\cdot10^{15}$~\neutron. 

Figure \ref{fig:DV} shows $\Delta$V$_{10fC}$($\Phi$) as a function of the acceptor removal coefficient \textit{c} [cm$^2$]. Green markers show $\Delta$V$_{10fC}$ ($\Phi$~=~\fluence{8}{14}), whereas blue markers are for $\Delta$V$_{10fC}$ ($\Phi$~=~\fluence{1.5}{15}).

\subsection{Noise}

\begin{figure}[h]
\centering
\includegraphics[width=\textwidth]{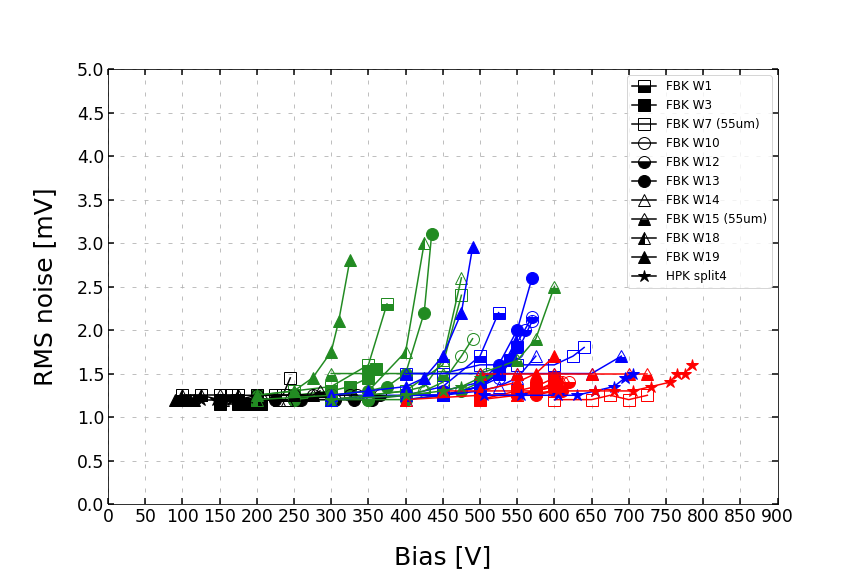}
\caption{RMS noise as a function of bias voltage. Sensors irradiated at different fluences are shown with different colors (black = pre-rad, green = $8\cdot10^{14}$~\neutron, blue = $1.5\cdot10^{15}$~\neutron, red =  $2.5\cdot10^{15}$~\neutron). Measurements performed at -25 $\degree$C.}
\label{fig:noise}
\end{figure}

The RMS noise as a function of the bias voltage is presented in figure~\ref{fig:noise}. The baseline noise of about 1.2~mV is due to the read-out board. For new sensors (black curves), the noise is always dominated by the read-out board.  In irradiated devices, the noise is higher than the baseline noise when the sensor has a large gain and therefore high leakage current. These conditions are met for sensors irradiated at $8\cdot10^{14}$~\neutron~and $1.5\cdot10^{15}$~\neutron; for sensors at $2.5\cdot10^{15}$~\neutron~ the noise increase is limited due to the small gain. 

In general, the noise is proportional to the squared sum of the leakage current $I_{leak}$, increasing linearly with the gain and logarithmically with the fluence, and the electronic noise of the read-out board\cite{vs_vertex2019}:
\begin{equation}
    Noise \propto \sqrt{ (Gain \cdot \ln(\Phi/\Phi_{0}))^2 + (Noise_{SC  board})^2}.
\end{equation}

where $\Phi_{0}$ is a constant with dimension [L$^{-2}$].

\begin{figure}[h] 
\centering 
\includegraphics[width=\textwidth]{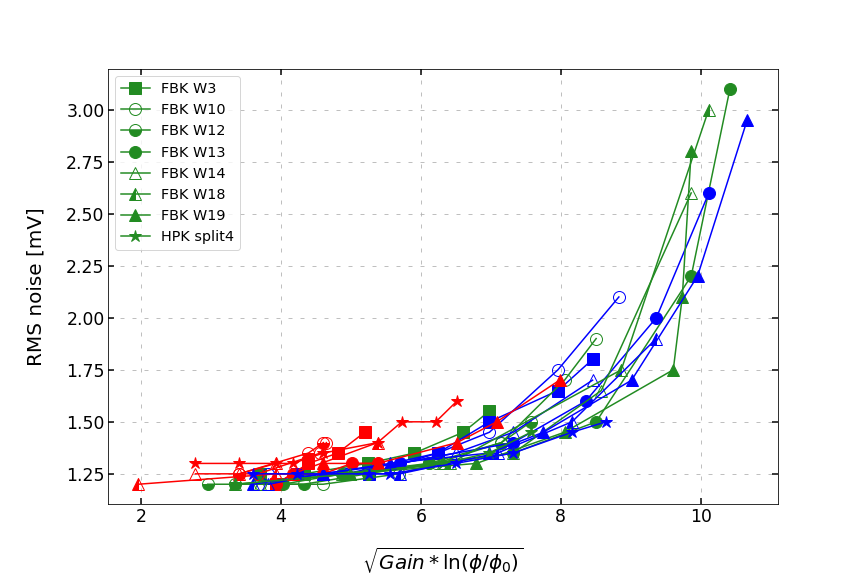}
\caption{RMS noise as a function of $\sqrt{Gain\cdot\ln{(fluence)}}$. A common  trend is observed for all irradiated sensors. Sensors irradiated at different fluences are shown with different colors (green = $8\cdot10^{14}$~\neutron, blue = $1.5\cdot10^{15}$~\neutron, red =  $2.5\cdot10^{15}$~\neutron). Measurements performed at -25 $\degree$C.}
\label{fig:noise_fluence}
\end{figure}

Figure \ref{fig:noise_fluence} shows the noise of irradiated sensors, plotted as a function of $\sqrt{Gain \cdot \ln{(\Phi/\Phi_{0})}}$. All tested sensors follow a common trend: this fact demonstrates that the noise is dominated  by the leakage current $I_{leak}$ and by the gain.
A noise level of 2-3~mV in irradiated devices does not spoil the time resolution, which remains below 40~ps, because such noise level is reached when the internal gain of the sensor is larger than 30 (see figure~\ref{fig:res-gain}); therefore, the signal-to-noise ratio remains large enough to achieve a good resolution. 

\section{Discussion of results: performance of un-irradiated sensors}

\subsection{Time resolution as a function of gain and electric field.}

A good time resolution is obtained when both the sensor internal gain and the electric field (or equivalently the bias voltage) at which such gain is reached are sufficiently high. None of these two conditions alone is enough. In fact, the jitter term of the time resolution is minimized~\footnote{In UFSD, the best time resolution is obtained when the jitter term is minimized~\cite{MF_book}.} when the leading-edge slew rate, $dV/dt$, is as steep as possible: (i) $dV$ depends upon the signal amplitude (hence the gain); (ii) $dt$ upon the electric field. 

For what concerns (ii), in UFSDs the majority of the signal is produced by holes, therefore it is important, from a sensor design point of view, that the holes drift velocity is saturated when the sensor is operated in the gain range 20-30 (the optimal one for UFSDs), so that $dt$ is minimized. The holes drift velocity requires a high electric field to approach saturation ( 50-100~kV/cm~\cite{MF_book}):  if the gain implant is too doped, the field in the bulk will not be able to reach high enough values before the sensor breakdown.

%A good time resolution is obtained when both the sensor internal gain and the electric field (or equivalently the bias voltage) at which such gain is reached are sufficiently high. None of these two conditions alone is enough. In fact, the jitter term of the time resolution is minimized when the leading-edge slew rate, $dV/dt$, is as steep as possible: $dV$ depends upon the signal amplitude (hence the gain), while $1/dt$ upon the electric field. 

%In UFSDs the majority of the signal is produced by holes, whose drift velocity is proportional to $dt$ and requires a high electric field to approach saturation. From a sensor design point of view, it is important that the holes drift velocity is saturated (electric field in the bulk $\geq$50~kV/cm) when the sensor is operated   in the gain range 20-30, so that $1/dt$ is minimized. If the gain implant is too doped, the field in the bulk will not be able to reach high enough values  before the sensor breakdown.

The pre-irradiated UFSD3.2 W19 is an example of a sensor with a too doped gain layer: large signals are reached at low bias, and the resolution is not as good as less doped sensors. 

\begin{figure}[hbtp]
\centering
    \includegraphics[width=0.9\linewidth]{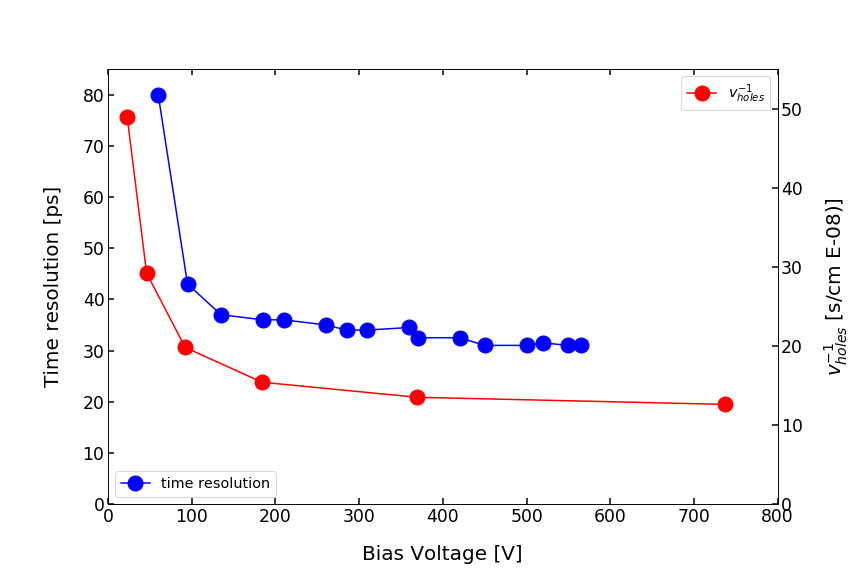}
    \caption{Time resolution (blue) as a function of the bias voltage at which the sensor reaches a gain of 20; each blue marker represents a different device. The red curve is the inverse of the holes drift velocity as a function of the bias voltage. Both the experimental results and the computation of the drift velocities are performed at a temperature of -25 $\degree$C.}
    \label{fig:res_velocity}
\end{figure}

Figure~\ref{fig:res_velocity} shows  the time resolution (blue) of different pre-rad sensors with gain = 20 (each marker represents a different device) as a function of the bias voltage, superposed with the inverse of the computed holes drift velocity. The plot well demonstrates that, for a fixed gain, sensors operated at higher bias voltage reach a better resolution because the holes drift velocity is higher.

\subsection{Time resolution as a function of the sensor thickness}
\label{sec:timeres-vs-thickness}

\begin{figure}[htbp]
    \centering
    \includegraphics[width=0.9\linewidth]{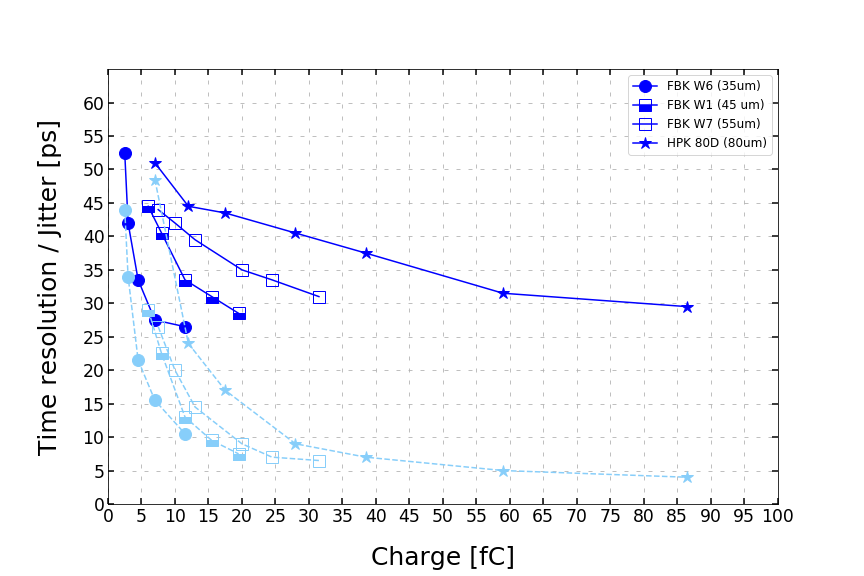}
    \caption{Time resolution as a function of the collected charge for sensors with four different active thicknesses. The dark blue curve represents the total resolution, whereas the light blue one is the jitter term only. Measurements performed at -25 $\degree$C.}
    \label{fig:res_4thick}
\end{figure}

Sensors with four different active thicknesses were compared in this study:  35, 45, 55\mum-thick sensors, belonging to the UFSD3.2 production, and 80\mum~thick sensors, manufactured by HPK. All measurements were performed at -25 $\degree$C. Figure \ref{fig:res_4thick} shows the time resolution  (dark blue) and the jitter term (light blue) as a function of the collected charge. The jitter term is computed analytically as the ratio between the noise and the signal slew rate ($dV/dt$):

\begin{equation}
   \sigma_{jitter} = Noise/(dV/dt)
\label{eq:jitter}   
\end{equation}

\newpage
 
\begin{figure}[htbp!]
    \centering
    \includegraphics[width=0.9\linewidth]{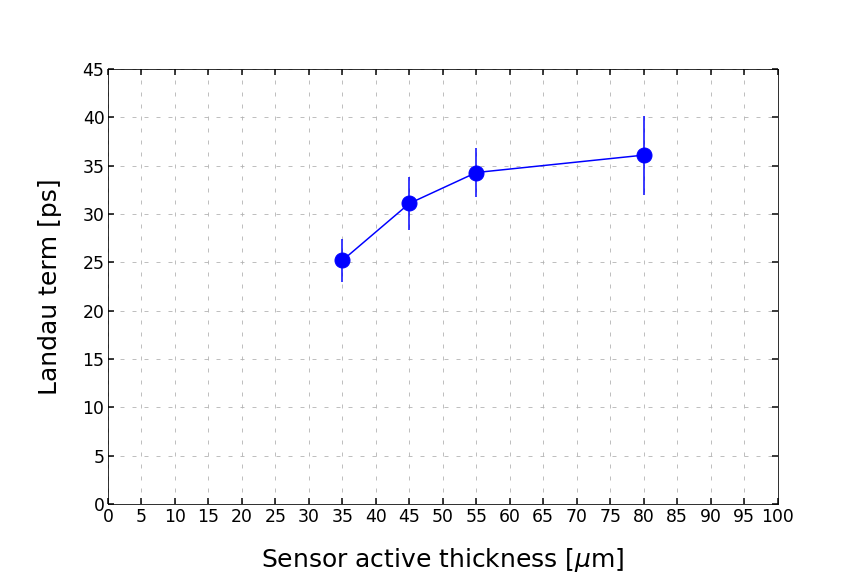}
    \caption{Landau term as a function of the sensor active thickness.}
\label{fig:landau_term}
\end{figure}

Two observations:
\begin{itemize}
    \item  In thicker sensors, a given value of jitter is obtained at higher charges. The jitter term can be written as:
    \begin{equation}
    \sigma_{jitter} = Noise/(dV/dt) \propto Noise/Gain     
    \end{equation}. 
    Assuming a constant noise value (the read-out board noise), and a given gain value, the collected charge scales with the sensor active thickness, and that explains the spread of the light blue curves in figure~\ref{fig:res_4thick}.
    \item The non-uniform energy deposition generated by an impinging MIP, amplified by the gain, creates variations of the signal shape on an event-to-event basis (the so-called Landau noise~\cite{ROPP}). The related uncertainty $\sigma_{Landau}$, which limits the time resolution, can be computed as the difference in quadrature of the total resolution and the jitter: 
    \begin{equation}
    \sigma_{Landau} = \sqrt{\sigma_{total}^{2} - \sigma_{jitter}^{2}}
    \end{equation}
    The Landau term as a function of the active thickness is reported in figure~\ref{fig:landau_term}: it decreases in thinner sensors, as expected~\cite{SADROZINSKI201618}.
\end{itemize}

\subsection{Time resolution as a function  
of the energy deposited in the event.}
%not sure about the title ...in the Landau distribution.}

The energy deposition of MIPs crossing a thin layer of silicon follows a Landau distribution. Events in the lower tail of the distribution tend to have low and uniform energy-per-unit-length deposits, while events in the upper tail have very large localized deposits. Therefore, the position of an event in the Landau distribution is an indicator of the uniformity of the energy deposition: the higher the energy deposited, the higher the non-uniformity. 

\begin{figure}[htbp!]  
\centering 
\includegraphics[width=0.9\textwidth]{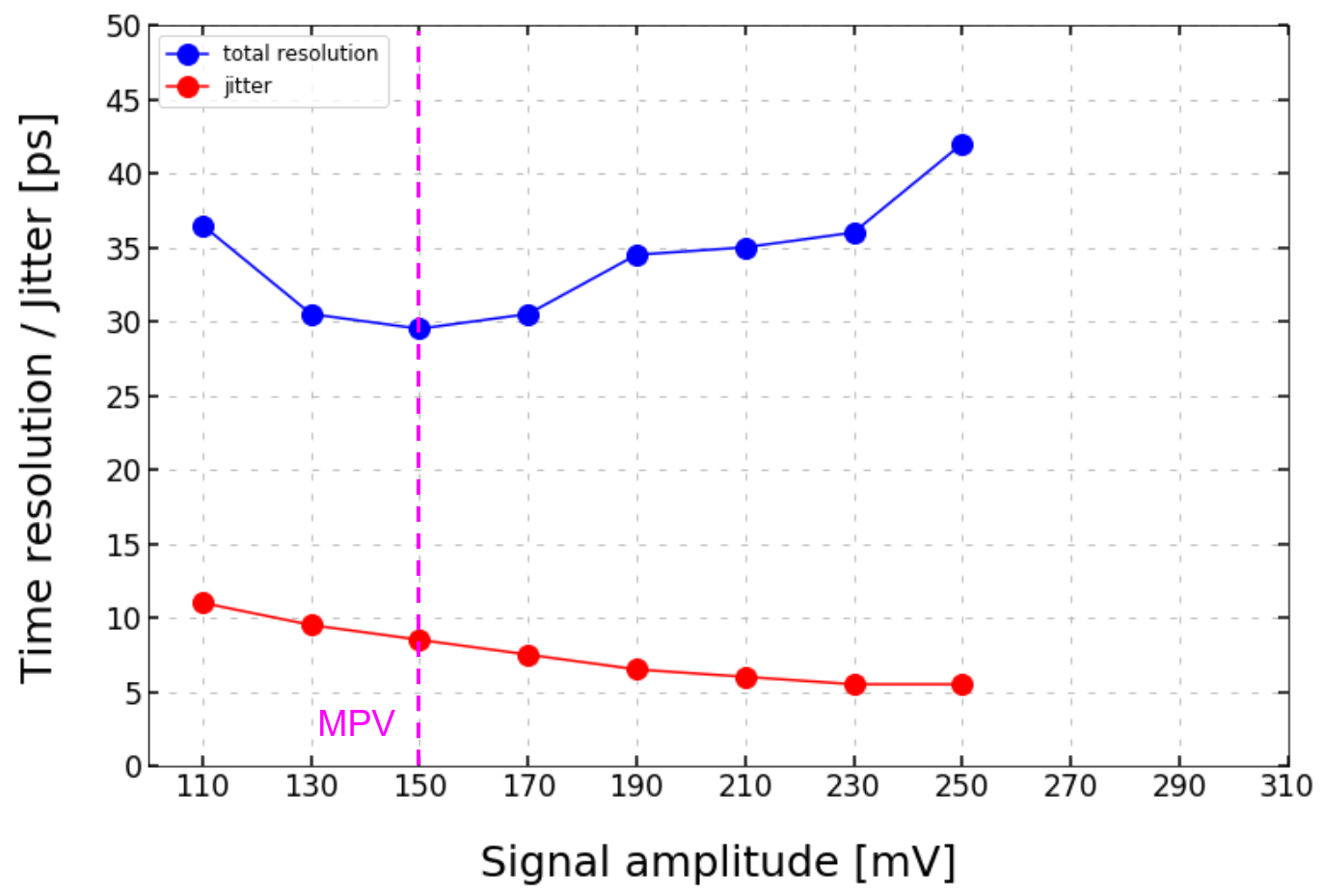}
\caption{Time resolution (blue) and jitter term (red) in bins of signal amplitude. The Landau MPV is $\sim$~150~mV, corresponding to the best time resolution. Measurements performed at -25 $\degree$C.}
\label{fig:slices_of_landau}
\end{figure}

Figure~\ref{fig:slices_of_landau} shows the time resolution and the jitter term (measured at -25 $\degree$C) in bins of the signal amplitude for a sensor from UFSD3.2 W7 (55\mum-thick). As expected, the jitter contribution decreases with amplitude. The time resolution, instead,  first improves as a function of signal amplitude, due to the smaller jitter contribution, then it worsens at high signal amplitudes due to the much larger Landau noise term. 

\begin{figure}[tb!]
\centering
\includegraphics[width=0.9\textwidth]{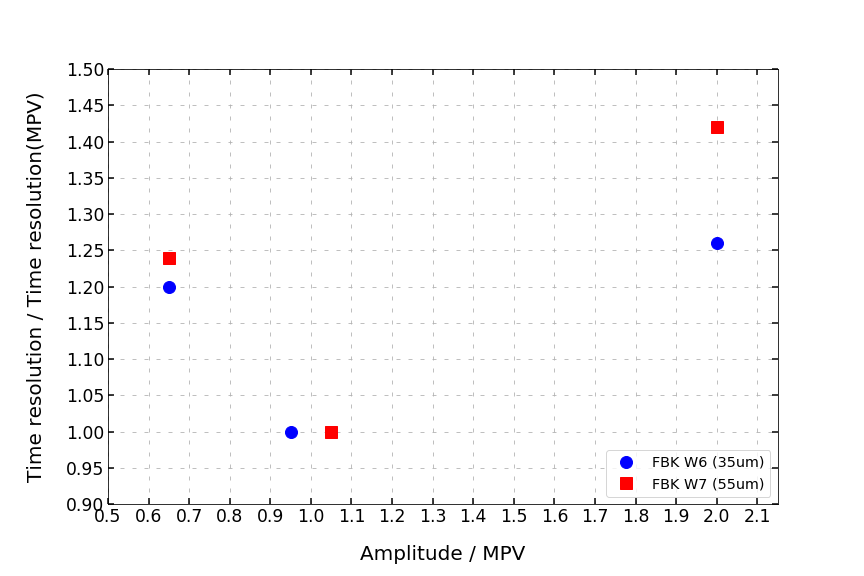}
\caption{Relative time resolution in 3 bins of amplitude (expressed in unit of the MPV) for UFSD3.2 W6 (35\mum~ thick) and W7 (55\mum~ thick). The markers in x=1 are slightly shifted only for representation. Measurements performed at -25 $\degree$C.}
\label{fig:w7_w6}
\end{figure}

A similar study has also been performed with UFSD3.2  W6 (35\mum~ thick) to investigate the effect of the sensor thickness. Since W6 is thinner, the worsening of the resolution in the upper tail of the distribution should be milder than for UFSD3.2 W7. Figure~\ref{fig:w7_w6} shows the time resolution, normalized to the resolution obtained at the MPV, for UFSD3.2 W6 and W7, in three bins of amplitude (expressed in units of  MPV). As expected, the worsening of the Landau term is larger in thicker sensors.

\section{Discussion of results: performance of irradiated sensors}

\begin{figure}[htb]
\centering
\includegraphics[width=\textwidth]{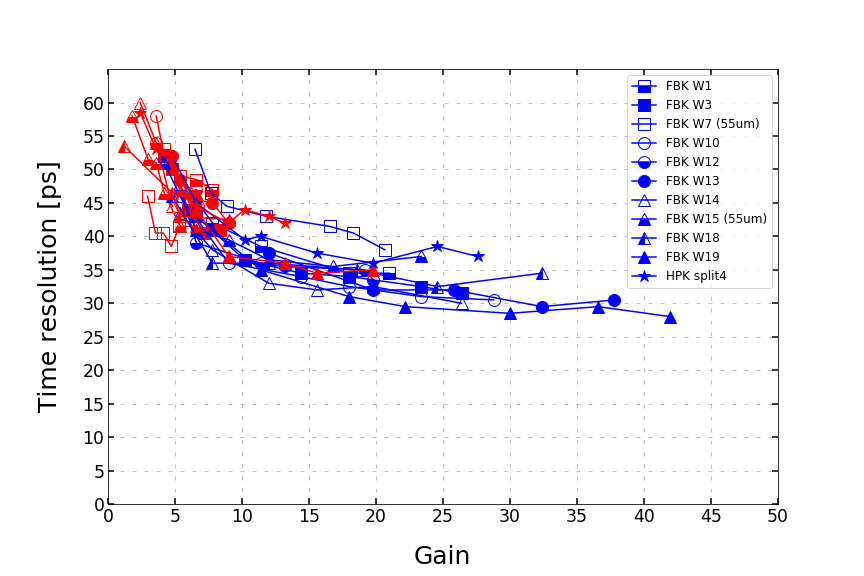}
\caption{Time resolution as a function of Gain. Sensors irradiated at different fluences are shown with different colors (black = pre-rad, green = $8\cdot10^{14}$~\neutron, blue = $1.5\cdot10^{15}$~\neutron, red =  $2.5\cdot10^{15}$~\neutron). Measurements performed at -25 $\degree$C.}
\label{fig:res-gain}
\end{figure}

Focusing on the time resolution of irradiated devices, it is important to notice that, above a certain bias voltage, once the holes drift velocity is saturated, the time resolution improves with gain in a very similar way for all the wafers. This common trend can be observed in figure~\ref{fig:res-gain}, representing the time resolution as a function of the gain for sensors irradiated at a fluence of ~$1.5\cdot10^{15}$~\neutron~and ~ $2.5\cdot10^{15}$~\neutron~(all biased above 300 V).

\subsection{Time resolution of sensors 45 or 55\mum~thick.}
%It can be noticed that the sensor thickness plays an important role.

\begin{figure*}[htbp]
\centering
\begin{subfigure}[b]{\textwidth}
\centering
\includegraphics[width=\textwidth]{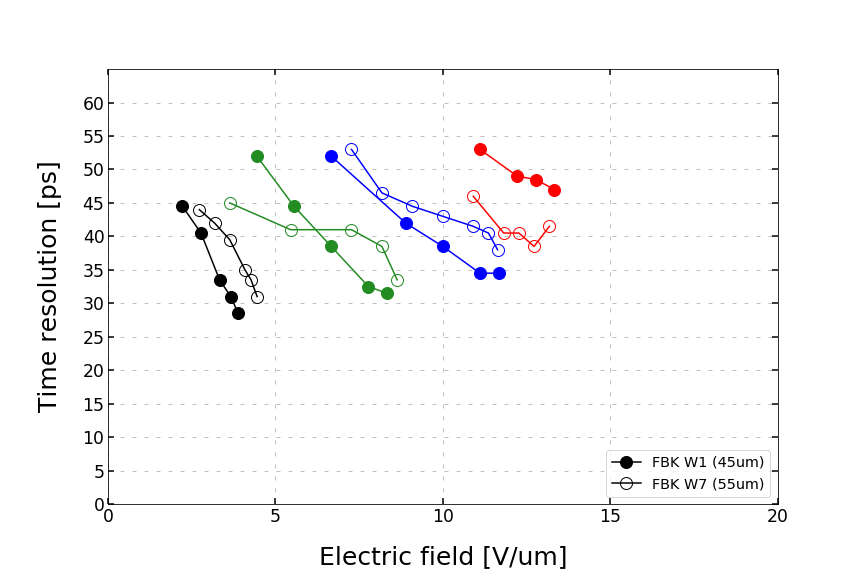}
\label{fig:el_field_shallow}
\end{subfigure}
\hfill
\begin{subfigure}[b]{\textwidth}  
\centering 
\includegraphics[width=\textwidth]{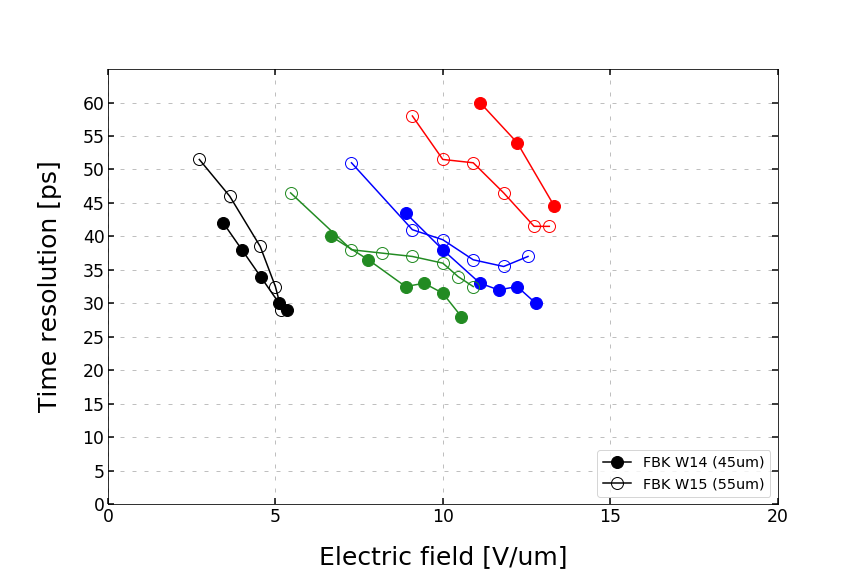}
\label{fig:el_field_deep}
\end{subfigure}
\caption{Time resolution as a function of the electric field in FBK UFSD3.2 sensors with a shallow (\textit{top}) or deep (\textit{bottom}) gain implant for two different active thickness (45 vs 55\mum). Measurements performed at -25 $\degree$C.}
\label{fig:el_field}
\end{figure*}

The UFSD3.2 production features pairs of wafers with the same gain layer design but different active thicknesses. One pair is formed by W1 (45 $\mu$m) and W7 (55 $\mu$m) and a second pair by W14 (45 $\mu$m) and W15 (55 $\mu$m). W1-W7 have a shallow gain implant, while W14-W15 a deep implant. Figure~\ref{fig:el_field} shows the time resolution of those two pairs, as a function of the  electric field in the sensor bulk, calculated as the ratio between V$_{bias}$ and the sensor active thickness.

Thin sensors have a better resolution at a given value of electric field (see section \ref{sec:timeres-vs-thickness}), as the Landau noise is larger in thicker devices. At high fluence, $2.5\cdot10^{15}$~\neutron, the sensor gain is close to 1: sensors with thicker bulk have an increased initial charge deposition, leading to a better time resolution. 

Hence, thinner sensors, with either shallow or deep gain implants, have better performance than thicker ones, if the sensor gain is high enough. Conversely, thicker sensors are more performing at low or no gain.

\subsection{Carbon co-implantation}

The beneficial effect of the co-implantation of carbon in the boron gain layer can be quantified using, as figure of merit, the voltage increase needed to provide 10~fC after a fluence $\Phi$,  $\Delta$V$_{10fC}$($\Phi$). 

Pre-rad HPK2 split 4 (not carbonated), UFSD3.2 W18, and W19 (both carbonated) sensors behave similarly and have similar gain layers designs. Nevertheless, their behavior differs with irradiation. After a fluence $8\cdot10^{14}$~\neutron, HPK2 split 4 $\Delta$V$_{10fC}$($\Phi$~=~\fluence{8}{14})
$\sim$~350~V, against $\sim$~200~V for UFSD3.2 W18 and W19 (see figure~\ref{fig:DV}). This difference increases further at higher fluences: at $1.5\cdot10^{15}$~\neutron, HPK2 split 4 $\Delta$V$_{10fC}$($\Phi$~=~\fluence{1.5}{15}) $\sim$~600~V while for UFSD3.2 W18, and W19 is $\sim$~350~V. 

%This difference is rather remarkable, and it has an important impact on the operation of the devices.

\subsection{Shallow or deep gain implant}

The position of the gain implant has consequences on the radiation hardness of the sensor design. 

\begin{figure}
    \centering
    \includegraphics[width=\linewidth]{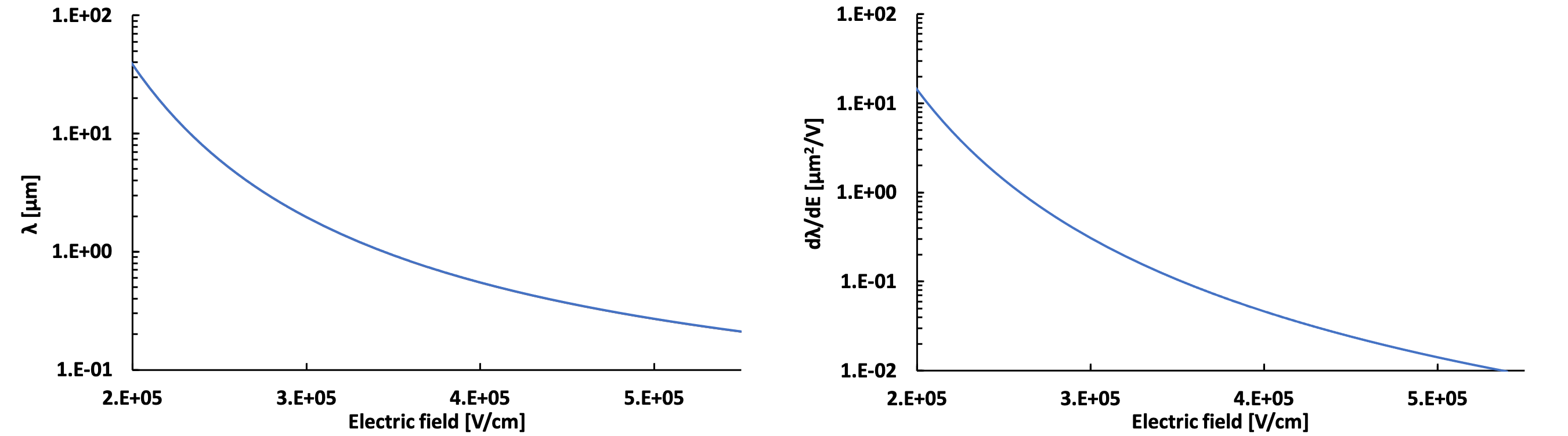}
    \caption{Electrons mean free path $\lambda$ between two subsequent scattering events producing secondary charges (\textit{left}) and $d\lambda/dE$ (\textit{right}) at 300~K as a function of the electric field $E$, according to the Massey impact ionization model~\cite{1677871}.}
    \label{fig:lambda}
\end{figure}

An important parameter governing the internal gain in UFSDs is the path necessary (on average) to acquire enough energy to achieve charge multiplication, $\lambda$. Indeed, the UFSD gain can be written as~\cite{CARTIGLIA-34thRD50}:

\begin{equation}
    Gain \propto e^{\frac{d}{\lambda}}
\label{eq_gain_2}
\end{equation}

where $d$ is the width of the gain layer, and $\lambda$ is a function of the electric field, as presented in~\ref{fig:lambda} (\textit{left}). A thorough explanation of this subject can be found in~\cite{CARTIGLIA-34thRD50}.

In deep implants, $d$ is wider than in shallow implants (see section \ref{sec:ufsd32}), and this leads to higher gain. Consequently, a deep implant has to be doped less, in order to increase $\lambda$ and generate the same gain as a shallow implant. Decreasing the gain layer doping in deep implants worsens their radiation resistance, since less doped implants are more affected by the acceptor removal mechanism~\cite{FERRERO201916}.  However, there is another important parameter that influences the radiation resistance of gain implants: the gain recovery capability with the external bias.

The value of the electric field in the gain region is usually $\sim400$~kV/cm ($\sim300$~kV/cm) in a shallow (deep) implant. With irradiation this electric field drops, increasing $\lambda$; the gain can be restored by increasing the electric field in the gain layer, that is by raising the bias voltage of the sensor.  The effectiveness of the recovery  depends on the value of the electric field and of $\lambda$: it is more effective in a deep gain layer, characterized by a lower electric field, since $d\lambda/dE$ is larger~\cite{CARTIGLIA-34thRD50}, as shown in~\ref{fig:lambda} (\textit{right}). 

For this reason, deep gain layer designs have a higher gain recovery capability with bias.

The interplay between intrinsic radiation resistance of the gain layer (acceptor removal coefficient), and the position of the gain implant (gain recovery capability),  can be illustrated by analyzing the properties of UFSD3.2 W3 (shallow implant) and W13 (deep implant). 

When new, they have roughly the same gain, as shown in figure~\ref{fig:charge}, and similar  values of the acceptor removal coefficient \textit{c} (see~\ref{fig:DV}). However, W3 has a higher value of $\Delta$V$_{10fC}$($\Phi$~=~\fluence{8}{14}) and $\Delta$V$_{10fC}$($\Phi$~=~\fluence{1.5}{15}), as shown in figure~\ref{fig:DV}, therefore it is less radiation resistant.

\subsection{Thermal treatments}

\begin{figure}[htb]
    \centering
    \includegraphics[width=\linewidth]{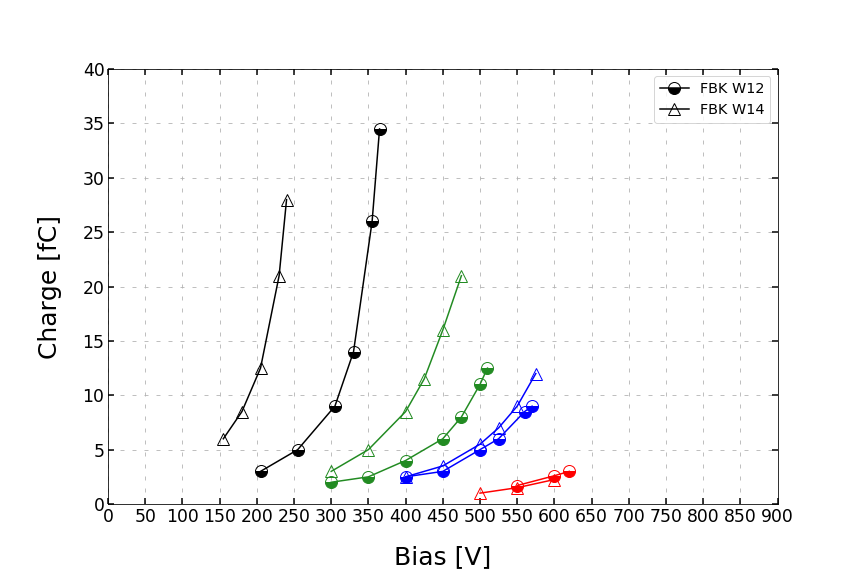}
    \caption{ UFSD 32 W14 (CBH) requires an higher increase of bias voltage to compensate for the effects of radiations than UFSD3.2 W12 (CBL). Sensors irradiated at different fluences are shown with different colors (black = pre-rad, green = $8\cdot10^{14}$~\neutron, blue = $1.5\cdot10^{15}$~\neutron, red =  $2.5\cdot10^{15}$~\neutron). Measurements performed at -25 $\degree$C.}
    \label{fig:w12-w14}
\end{figure}

One of the steps in the sensor production is the activation of the gain implant dopants. In the UFSD3.2 production, this effect has been studied by using two different procedures to activate the carbonated deep gain implants. The two different sequences are called Carbon-Boron Low (CBL, W10,W12 and W13) and Carbon-Boron High (CBH, W14, W15, W18, and W19), where low and high refer to the thermal load. Overall, they have similar performance with one notable difference: CBL is more radiation resistant~\cite{MF_book}.  Figure \ref{fig:w12-w14} illustrates this fact: UFSD3.2 W14 (CBH) has a higher pre-rad gain than UFSD3.2 W12, but, with increasing irradiation, the difference gets smaller and, at $1.5\cdot10^{15}$~\neutron, they operate at the same voltage. In other words, UFSD3.2 W12 has a smaller $\Delta$V$_{10fC}$($\Phi$).

\section{Conclusions}

In this work, the performance of several UFSDs have been measured using a telescope instrumented with a $^{90}$Sr $\beta$-source.
Different gain layer designs have been studied:  shallow and deep implants, with or without carbon co-implantation, activated with low or high thermal load.  The performance of these different designs are reported in terms of: the collected charge $Q$, the noise, the steepness $\alpha$ of the $Q(V)$ curve, the bias to deliver 10~fC V$_{10fC}$, the bias increase $\Delta$V$_{10fC}$($\Phi$) to deliver 10~fC after a fluence $\Phi$, and the time resolution $\sigma_t$. 

 The most important outcomes of this work are:

\begin{itemize}
    \item The introduction of $\Delta$V$_{10fC}$($\Phi$), which proved to be an effective figure to assess the radiation resistance and the sensitivity to non-uniform irradiation of a gain layer design: the smaller $\Delta$V$_{10fC}$, the higher the radiation hardness and the lower the sensitivity to non-uniform irradiation. 
    \item Carbon co-implantation decreases $\Delta$V$_{10fC}$($\Phi$) by about 50\% at every fluence.
    \item Radiation resistance depends upon two parameters: (i) the acceptor removal coefficient \textit{c} and (ii) the bias recovery capability, quantified by $\Delta$V$_{10fC}$($\Phi$). The interplay of these two aspects determines that deep gain implants, despite having a slightly larger acceptor removal coefficient, are more radiation resistant than shallow gain implants.
    \item All sensors were able to reach a time resolution of 30 to 40~ps up to a fluence of $1.5\cdot10^{15}$~\neutron, with at least 10~fC of charge delivered. At $\Phi = 2.5\cdot10^{15}$~\neutron~the signal charge decreases below 10~fC and the resolution worsens.
    \item The best time resolution is obtained when the electric field in the bulk is large enough: high gain at low bias voltage leads to poor performances.  
    \item $Q(V)$ characteristics with small $\alpha$ values are to be preferred since they lead to: (i) a better $\sigma_t$ given the higher V$_{10fC}$, and a decrease of the effects of either (ii) doping or (iii) bias non-uniformity. 
    \item Pre-rad low values of V$_{10fC}$ provide a larger possible bias increase; however, they also imply very steep $Q(V)$ and a worse $\sigma_t$. 
\end{itemize}

Thanks to these results, the FBK UFSD3.2 carbonated deep implant with CBL thermal diffusion scheme (W13) has been identified as the best design, able to achieve 40~ps up to a radiation fluence of~\fluence{2.5}{15} delivering at least 5~fC of charge. This design is sufficiently doped to limit the power consumption when new, but it also features a rather shallow charge-bias characteristic, ensuring a low sensitivity to non-uniform biasing conditions, and the operation at a sufficiently high bias voltage. W13 features one of the lowest $\Delta$V$_{10fC}$($\Phi$), proving to be highly resistant to radiations and not particularly susceptible to non-uniform irradiation. 

This study also analyzes the performance of UFSDs with different active thicknesses (35, 45, 55, and 80\mum-thick):

\begin{itemize}
    \item Thin devices have a lower Landau noise term.
    \item At high radiation levels, $\Phi = 2.5\cdot10^{15}$~\neutron~, when the gain value is very small, thicker sensors have better time resolution given their higher initial signal.
    \item The time resolution has been studied in bins of the Landau distribution. The best results are achieved around the most-probable-value. For smaller amplitudes, the jitter term dominates, while for higher values, the Landau noise is larger,  worsening the time resolution. 
\end{itemize}

Finally, it has been shown that deep CBL sensors have a lower $\Delta$V$_{10fC}$($\Phi$) than their CBH counterpart, therefore they should be preferred; in addition, a study on the dependence of the noise upon the radiation fluence has been performed: the sensor noise increases with the logarithm of the fluence, and it is mostly independent from the gain layer design. 

\section{Acknowledgements}
We kindly acknowledge the following funding agencies and collaborations: INFN – FBK agreement on sensor production; Dipartimenti di Eccellenza, Univ. of Torino (ex L. 232/2016, art. 1, cc. 314, 337); Ministero della Ricerca, Italia, PRIN 2017, Grant 2017L2XKTJ – 4DinSiDe; Ministero della Ricerca, Italia, FARE,    Grant R165xr8frt\_fare

\end{document}